\documentclass[ aip,
 jmp, 
 reprint,]{revtex4-2}
\usepackage{eurosym}
\usepackage{amssymb}
\usepackage{graphicx}
\usepackage{dcolumn}
\usepackage{bm}
\usepackage{amsmath}
\usepackage{graphicx}
\usepackage{dcolumn}
\usepackage{bm}
\usepackage[utf8]{inputenc}
\usepackage[T1]{fontenc}
\usepackage{mathptmx}
\usepackage{natbib}
\usepackage{caption}
\usepackage{subcaption}
\usepackage{xcolor}
\usepackage[title]{appendix}

\begin{document}

\title{Ion acoustic and spin electron acoustic cnoidal waves in a spin polarized plasma with exchange effects}
\author{B. Sania}
\affiliation{Department of Physics, Lahore College for Women University,
Lahore, 54000, Pakistan.}

\affiliation{Department of Physics, Government
College University, Lahore, 54000, Pakistan.}

\author{Z. Iqbal}
\email{dr.zafariqbal@gcu.edu.pk}
\affiliation{Department of Physics, Government
College University, Lahore, 54000, Pakistan.}

\author{Ch. Rozina}
\affiliation{Department of Physics, Govt Gulberg College for Women, Gulberg Lahore, 54000, Pakistan.}

\author{Hafeez ur Rehman}
\affiliation{Theoretical Physics Division (TPD), PINSTECH, P.O. Nilore, Islamabad
45650, Pakistan}

\author{G. Abbas}
\affiliation{Department of Physics, Government
College University, Lahore, 54000, Pakistan.}

\date{\today }

\begin{abstract}Separate spin evolution-quantum hydrodynamic (SSE-QHD) model is employed to
address the nonlinear propagation of ion-acoustic wave (IAW) and spin
electron-acoustic wave (SEAW) in a spin polarized electron-ion plasma. The
analysis has been made under the self-consistent field approximation and with
exchange correlation effects. The reductive perturbation method (RPM) is used
to derive KdV equation and its cnoidal wave solutions. We noted that the phase
velocity of IAW in the self-consistent field approximation is almost constant
however, in the presence of exchange-correlation potential there is an abrupt
change in the phase velocity. The phase velocity of SEAW decreases in the
presence of exchange-correlation effects as compare to self-consistent field
approximation. We have calculated the condition for the existence of
\ nonlinear structures and it is found that \ in the presence of exchange
effect the condition satisfy for certain values of $\eta$ at different
densities. Furthermore, the comparisons have been made with and without
exchange effects, it shows that although the nonlinear profiles of both waves
are significantly\ affected with exchange effect but it also converts cnoidal
structures of SEAW from rarefactive to compressive. The influence of
exchange-correlation potential and spin polarization on the \ profiles of both
nonlinear structures is evaluated numerically. The present study may be
helpful to understand formation of \ new longitudinal cniondal structures in
laboratory degenerate plasma.
\end{abstract}

\maketitle

\preprint{AIP/123-QED}

% Force line breaks with \\

%Lines break automatically or can be forced with \\

\section{Introduction}

The study of cnoidal waves, an important extension of nonlinear structures,
has gained considerable attention due to their wide applications in nonlinear
transport phenomena. These waves represent a significant class of nonlinear,
periodic solutions derived from the Korteweg-de Vries equation.
Mathematically, they are defined using the Jacobi elliptic cn function, which
gives them the name "cnoidal waves." Numerous researchers have extensively
explored this intriguing phenomenon \cite{Shamy 11,rehman 12,S. Mahmood,5 ref
for ion} to investigate the unique properties and behaviors of cnoidal waves
in various plasma environments.

The nonlinear uniform signals observed at the edge of the Tokamak can be
described by cnoidal waves \cite{8}. Shamy \cite{Shamy 11} examined the
behavior of ion-acoustic (IA) cnoidal waves, which involved solving the
nonlinear equations for both cnoidal and solitary waves, leading to the
formation of compressive structures in both cases. By employing RPM, the
propagation of both linear and nonlinear fast magnetohydrodynamic cnoidal
waves in electron-positron (e-p) plasmas was studied in \cite{rehman 12}.
Soliton solutions were found, and the influence of plasma beta on wave
dispersion was analyzed. Mahmood and Haas investigated ion-acoustic cnoidal
waves \cite{S. Mahmood} and solitons in unmagnetized quantum plasmas,
resulting in the formation of compressive and rarefactive nonlinear cnoidal
structures in a degenerate plasma. R. Kaur et al. \cite{5 ref for ion} studied
IA cnoidal waves in a magnetized quantum plasma, considering the presence of
inertial positive ions, weakly relativistic ion beams, and trapped electrons,
and found that only positive potential IA cnoidal waves formed. The behavior
of acoustic cnoidal waves in pair-ion plasmas, involving ions of the same mass
but different temperatures, was explored in Ref \cite{14}, revealing
temperature-dependent compressive and rarefactive cnoidal wave structures. The
effect of positron density on the propagation of IA cnoidal waves in
electron-positron-ion plasmas was also investigated in \cite{15}. The QHD
model was generalized with contribution of electrons's spin effect
\cite{Kuzmenkov,Brodin-2007,Pavel-E-2015}. Using the QHD model for spin-1/2
particles it has been investigated that the spin effects not only modify the
dispersive properties of waves and instabilities in degenerate quantum plasmas
but also gives birth to new spin dependent linear and nonlinear waves
\cite{Mushtaq-12,Shahid-15,Zafar-1-17,Zafar-2-17,Gul-2020,Zafar1-18,Ayub-19,Zafar19}%
. The QHD model for spin-1/2 particles in which electrons of spin-up and
spin-down treated as a different fluid is called separate spin
evolution-quantum hydrodynamic (SSE-QHD) model. The significant application of
this model is a prediction of existence of new longitudinal and transverse
waves in degenerate plasma. One of new longitudinal wave i.e., spin electron
acoustic (SEA) wave can be used to explain the mechanism of high-temperature
superconductivity. This mechanism allows us to expect the superconductivity at
room temperatures if the electron's concentration would be high enough
\cite{andreev5}. Further, SSE-QHD equations have been employed to investigate
the solitary structures in the electron-ion
\cite{Zafar-PLA,Zafar-CTP,Pavel-2016} and electron-positron-ion
plasma\cite{Zafar-16}.

For dense plasmas, the electron interactions \cite{exc} are governed by the
short range electrostatic potential arising due to\ the Hartree term and the
electron exchange-correlation potential appears due to the overlying of wave
functions of dense fermions \cite{pavel 4,17,my2}, both exchange and
correlation effects of electrons are function of Fermi density under local
density approximation \cite{18,19}. Crouseilles et al. \cite{N. Crouseilles 3}
were the pioneers to introduce the exchange-correlation potential into the QHD equations.

Exchange effects in quantum plasmas have long been a subject of study. Various
methods and results have been developed in this area, but here we focus on
examining the exchange-correlation effects in spin-polarized plasma using the
SSE-QHD model \cite{Pavel-2016}. To the best of our knowledge, the periodic
wave structures of ion-acoustic (IA) and spin electron acoustic (SEA) waves,
influenced by exchange interactions in spin-polarized plasma, have not been
explored previously. In this work, we compare the characteristics of periodic
IA and SEA waves in a magnetized spin-1/2 quantum plasma, considering both the
self-consistent field approximation and exchange effects. Our findings show
that exchange effects play a key role in the transition from rarefactive to
compressive periodic wave structures. The manuscript is organized as follows:
Section 2 presents the mathematical model for the propagation of nonlinear
ion-acoustic waves in polarized plasmas. In Section 3, the nonlinear
Korteweg-de Vries (KdV) equation is derived using the reductive perturbation
method. Section 4 discusses the periodic solution of the KdV equation through
the Sagdeev potential approach. Section 5 presents numerical plots of IA and
SEA cnoidal waves, and Section 6 summarizes the key conclusions of the research.

\section{Basic Equations}

A partially spin-polarized, collisionless electron-ion plasma is examined to
study the one-dimensional non-linear IA wave in a degenerate, magnetized
plasma. In this scenario, we assume that the wave propagates parallel to the
external magnetic field $B_{0}||$ $\widehat{z}.$ Additionally, we neglect the
quantum effects related to the ions due to their large inertia, treating them
instead as a classical fluid governed by hydrodynamic equations:%
\begin{equation}
\partial_{t}n_{i}+\nabla.(n_{i}\mathbf{v}_{i})=0, \label{1}%
\end{equation}%
\begin{equation}
m_{i}n_{i}(\partial_{t}+\mathbf{v}_{i}.\nabla)\mathbf{v}_{i}=q_{i}%
n_{i}\mathbf{E,} \label{2}%
\end{equation}
where $n_{i}$ and $\mathbf{v}_{i}$ are the density and the velocity of plasma
ions, $q_{i}=e$ \ is a charge , and $m_{i}$ is the mass of ions respectively.
The electrons consist of two components i.e., one is spin-up and the other is
spin-down. In order to deal with electron dynamics we employ the SSE-QHD
model, which is modified with exchange correlation effects\cite{Pavel-2016}.
In this model, it is assumed that these two types of electrons are treated as
a separate fluid. Therefore, the continuity equation for each species is given
as;%
\begin{equation}
\partial_{t}n_{es}+\nabla.(n_{es}\mathbf{v}_{es})=(-1)^{l+1}\frac{\gamma
}{\hbar}(B_{y}S_{ex}-B_{x}S_{ey}), \label{3}%
\end{equation}
where $s=u,d$ for the spin-up and spin-down state of particles, $n_{es}$ and
$\mathbf{v}_{es}$ are the concentration and velocity field of electrons being
in the spin state $s$, $l=1$, for spin-up and $l=2$ for spin-down, $S_{ex}$
and $S_{ey}$ are the spin density projections each describing the evolution of
the spin-up and spin-down states of particles. The right hand side of
Eq.\ref{3} is normally zero but for the\ case SSE it is not equal to zero
which describe the non conservation of spin-up and spin-down particles.
However, if we treat the electrons as a single fluid, the usual form of
continuity is retained. The momentum equation for electrons is given as;%
\begin{align}
&  \left.  m_{e}n_{es}(\partial_{t}+\mathbf{v}_{es}.\nabla)\mathbf{v}%
_{es}+\nabla P_{es}-\frac{\hbar^{2}}{4m_{e}}n_{es}\mathbf{\nabla}\left(
\frac{\Delta n_{es}}{n_{es}}-\frac{(\nabla n_{es})^{2}}{2n_{es}^{2}}\right)
\right. \nonumber\\
&  \text{ \ \ \ \ \ \ \ }\left.  =-en_{es}\text{ }\left(  \mathbf{E}+\frac
{1}{c}[\mathbf{v}_{es},\mathbf{B}]\right)  +(-1)^{(l+1)}\gamma_{e}n_{es}\nabla
B_{z}\right. \nonumber\\
&  \text{ \ \ \ \ \ \ \ \ \ \ }\left.  +\frac{\gamma_{e}}{2}(S_{ex}\nabla
B_{x}+S_{ey}\nabla B_{y})+(-1)^{(l+1)}m(\widetilde{\mathbf{T}}_{ez}%
-\mathbf{v}_{es}T_{ez})+\mathbf{F}_{Exe,dd},\right.  \label{4}%
\end{align}
where the second term on left hand side $P_{es}=(6\pi^{2})^{2/3}n_{es}%
^{5/3}\hbar^{2}/5m$ is the pressure of electrons, third term gives the
contribution of Bohm Potential, first term on RHS is the Lorentz force, second
and third terms are the action of x, y and z projections of magnetic field on
the magnetic moment (spin),$\ \widetilde{\mathbf{T}}_{ez}=\frac{\gamma_{e}%
}{\hbar}(\mathbf{J}_{(M)ex}B_{y}-\mathbf{J}_{(M)ey}B_{x})$, which is the
torque current, $T_{ez}=\frac{\gamma}{\hbar}(B_{x}S_{ey}-B_{y}S_{ex})$ where
$\mathbf{J}_{(M)ex}=(\mathbf{v}_{eu}+\mathbf{v}_{ed})S_{ex}/2-\hbar
S_{ey}\left(  \nabla n_{eu}/n_{eu}-\nabla n_{ed}/n_{ed}\right)  /4m_{e}$, and
$\mathbf{J}_{(M)ey}=(\mathbf{v}_{eu}+\mathbf{v}_{ed})S_{ey}/2+\hbar
S_{ex}\left(  \nabla n_{eu}/n_{eu}-\nabla n_{ed}/n_{ed}\right)  /4m_{e}$ are
represent spin currents. The force field in terms of the spin-down electron
concentration can be written as,%
\begin{equation}
\mathbf{F}_{Exe,dd}=\chi e^{2}(n_{d})^{\frac{1}{3}}\nabla n_{d}, \label{5}%
\end{equation}
where%
\begin{equation}
\chi=2^{\frac{4}{3}}(\frac{3}{\pi})^{\frac{1}{3}}\left(  1-\frac
{(1-\eta)^{\frac{4}{3}}}{(1+\eta)^{\frac{4}{3}}}\right)  . \label{6}%
\end{equation}
Here $\eta$ is the spin polarization $\eta=n_{0eu}-n_{0ed}$ $=n_{i0}(3\mu
_{B}B_{0}/2\varepsilon_{F_{e}})$, $\varepsilon_{F_{e}}$ is the Fermi energy of
the electrons. The concentrations of spin-up and spin-down electrons in terms
of equilibrium number density and spin polarization can be expressed as
$n_{0eu}=n_{0e}(1-\eta)/2$ and $n_{0ed}=n_{0e}(1+\eta)/2$. In case of
longitudinal propagating waves the spin evolution equations do not contribute
therefore we dot not mention here but their explicit form is given in
\cite{Pavel-E-2015}. It is worth mention here we consider that the wave is
propagating parallel to the magnetic field which is along $z$ axis therefore,
Lorentz force equals to zero. However, the magnetic field contribution \ comes
through the spin polarization of electrons, which affects the equation of
state and the force of the Coulomb exchange interaction. For the propagation
of longitudinal waves in electron-ion plasma, the required Poisson equation
can be written as,%
\begin{equation}
\mathbf{\nabla}.\mathbf{E}=4\pi e(n_{i}-n_{eu}-n_{ed}), \label{7}%
\end{equation}
The initial plasma state is neutral: $n_{0u}+n_{0d}=n_{e0}=n_{i0}$, where
$n_{e0}$ and $n_{i0}$ are the initial densities of the electrons and ions.

\section{Derivation of nonlinear KdV equation}

We apply the reductive perturbation method to derive the Korteweg--de Vries
(KdV) equation. To analyze the nonlinear ion-acoustic (IA) wave, we assume
that the electrons are inertialess. For one-dimensional propagation along the
z-direction, we introduce a scaling of the independent variables using the
following stretched coordinates:%
\begin{equation}
\xi=\epsilon^{\frac{1}{2}}\left(  z-Vt\right)  ,\text{ \ \ \ \ \ \ }%
\tau=\epsilon^{\frac{3}{2}}t, \label{8}%
\end{equation}
where $\epsilon$ is a small parameter measuring the strength of perturbation
or weakness of nonlinearity and $V$ is phase velocity of wave. The perturbed
quantities $n,v,$ and $\phi$ are expanded about their equilibrium position in
the form of small parameter $\epsilon$ as%
\begin{align}
n_{es}  &  =n_{e0s}+\epsilon n_{e1s}+\epsilon^{2}n_{e1s}+\cdots,\nonumber\\
n_{i}  &  =n_{i0}+\epsilon n_{i1}+\epsilon^{2}n_{i2}+\cdots,\nonumber\\
v_{esz}  &  =\epsilon v_{esz1}+\epsilon^{2}v_{esz2}+\cdots,\nonumber\\
v_{iz}  &  =\epsilon v_{iz1}+\epsilon^{2}v_{iz2}+\cdots,\nonumber\\
\phi &  =\epsilon\phi_{1}+\epsilon^{3}\phi_{2}+\cdots. \label{8b}%
\end{align}
Next, by utilizing the relationship $E=-\nabla\phi$ and expanding the
variables as defined above and collecting terms of lowest order in $\epsilon
$,\ from the continuity equations and equations of motion, we obtain the
perturbed number densities of electrons and ions in term of electric potential
$\phi_{1}$ as follows;%

\begin{equation}
n_{1es}=%
%TCIMACRO{\dsum \limits_{s=u,d}}%
%BeginExpansion
{\displaystyle\sum\limits_{s=u,d}}
%EndExpansion
\frac{n_{0es}\frac{e}{m}}{u_{Fes}^{2}}\phi_{1},\text{ \ \ \ \ \ }n_{1i}%
=\frac{n_{i0}\frac{e}{m_{i}}}{V^{2}}\phi_{1.} \label{9}%
\end{equation}
Here $u_{Fu}^{2}=(6\pi^{2})^{2/3}n_{0eu}^{2/3}\hbar^{2}/3m$ and $u_{Fd}%
^{2}=(6\pi^{2})^{2/3}n_{0ed}^{2/3}\hbar^{2}/3m-\chi e^{2}n_{0ed}^{1/3}$\ $/m.$
Substituting the above perturbed number densities in the Poisson's equation in
the lowest order of $\epsilon$ which is given as $n_{eu1}+n_{ed1}-n_{i1}=0,$
we get the expression for the phase velocity of IA wave in the normalized form
as%
\begin{equation}
w=\sqrt{\frac{\lambda}{\frac{3(1-\eta)^{\frac{1}{3}}}{2}+\frac{(1+\eta
)/2}{\left(  \frac{1}{3}(1+\eta)^{\frac{2}{3}}-\frac{\chi e^{2}n_{0e}^{1/3}%
}{mv_{Fe}^{2}}(\frac{1+\eta}{2})^{\frac{1}{3}}\right)  }}}, \label{10}%
\end{equation}
where we have used $w=V/v_{Fe},$ $\lambda=m_{e}/m_{i}$. The equation\ref{10}
shows the dependence of phase velocity on spin polarization and exchange
potential. The Fig.(1) shows the phase velocity of IA wave plotted against the
spin polarization $\eta.$ It is noted from the figure that in the
self-consistent field approximation the phase velocity of IA wave is almost
remains constant but slight increase in the phase velocity is notes near
$\eta\rightarrow$ $1$. Therefore, phase velocity for all the values of $\eta$
is positive. However, in the presence of exchange effects the drastic change
in the phase velocity is noted against $\eta$. For instance, it becomes
imaginary at certain points from $0.01$ to $0.2$. It means that to investigate
the non-linear cnoidal waves the allowed range of $\eta>0.2$.

Collecting next higher order of $\epsilon$ power and with the aid of
quantities of lowest order in $\epsilon$ , we get the Korteweg-de Vries (KdV)
equation as%
\begin{equation}
A\partial_{T}\phi_{1}+R\partial_{\xi}\phi_{1}^{2}+\partial_{\xi}^{3}\phi
_{1}=0,\label{11}%
\end{equation}
where the coefficients $A$ and $R$ are given as%
\begin{equation}
A=\frac{2A_{0}}{v_{Fe}\lambda_{De}^{2}},\text{ \ \ \ where \ \ }A_{0}%
=\frac{2\lambda}{w^{3}}\label{12}%
\end{equation}
and%
\begin{equation}
R=\frac{e}{mv_{Fe}^{2}\lambda_{De}^{2}}R_{0}\label{13}%
\end{equation}
where%
\begin{align}
R_{0} &  =\frac{3(\lambda)^{2}}{w^{4}}+\frac{1}{2(1-\eta)^{\frac{1}{3}}%
}\nonumber\\
+ &  \frac{\frac{1}{2}(1+\eta)}{(\frac{1}{3}(1+\eta)^{\frac{2}{3}}-\frac{\chi
e^{2}n_{0e}^{1/3}}{mv_{Fe}^{2}}(\frac{1+\eta}{2})^{\frac{1}{3}})^{2}}\left(
1-\frac{\frac{\chi e^{2}n_{0e}^{1/3}}{mv_{Fe}^{2}}(\frac{1+\eta}{2})^{\frac
{1}{3}}}{(\frac{1}{3}(1+\eta)^{\frac{2}{3}}-\frac{\chi e^{2}n_{0e}^{1/3}%
}{mv_{Fe}^{2}}(\frac{1+\eta}{2})^{\frac{1}{3}})}\right)  ,\label{13a}%
\end{align}
in the above equations, we have used $\lambda_{De}=v_{Fe}/\omega_{pe}.$

\section{Periodic wave solution of the KdV equation}

To derive the periodic wave solution of the KdV Eq.\ref{11}, we use the
standard transformation as: $\zeta=\xi-U_{0}\tau$ \ where $U_{0}$ is the speed
of nonlinear waves. This transformation will provide the following equation:%
\begin{equation}
-U_{0}\frac{d\phi_{1}}{d\zeta}+\frac{C}{2}\frac{d\phi_{1}^{2}}{d\zeta}%
+D\frac{d^{3}\phi_{1}}{d\zeta^{3}}=0, \label{14}%
\end{equation}
where $C=2R/A$ and $D=1/A$. After some algebraic simplifications, the equation
for Sagdeev pseudopotential $U(\phi_{1})$ is obtained as%
\begin{equation}
U(\phi_{1})=\frac{C}{6D}\phi_{1}^{3}-\frac{U_{0}}{2D}\phi_{1}^{2}+\rho_{0}%
\phi_{1}, \label{15}%
\end{equation}
where $\rho_{0}$ is the constant of integration. After some mathematical
calculations, we obtain $\frac{d\phi_{1}}{d\zeta}$ which is the phase plane
given as%
\begin{equation}
\frac{d\phi_{1}}{d\zeta}=\pm\sqrt{\frac{C}{3D}\left(  \phi_{0}-\phi
_{1}\right)  \left(  \phi_{1}-\phi_{2}\right)  \left(  \phi_{1}-\phi
_{3}\right)  }, \label{16}%
\end{equation}
where $\phi_{2}$ and $\phi_{3}$ are the real roots of Sagdeev potential. Here,
we have two equilibrium states $j_{\left(  1,2\right)  }$ named as center and
saddle point and are written as under;%
\[
j_{1,2}=\frac{U_{0}}{C}\pm2\sqrt{\frac{U_{0}^{2}}{C^{2}}-\frac{2D\rho_{0}}%
{C},}%
\]
where%
\begin{equation}
\Upsilon=\frac{U_{0}^{2}}{C^{2}}-\frac{2D\rho_{0}}{C}. \label{17}%
\end{equation}
In order to have a real values of potential well which is necessary for the
existence of cnoidal waves, the value of $\Upsilon>0$. Fig.(2) shows that
value of $\Upsilon\,$depends on the spin polarization and exchange effects.
For instance, as we increases the number density the allowed range of $\eta$
for the formation of cnoidal structures is decreases. It can be seen from the
Fig.(2), that at number density $n_{0}=10^{21}cm^{-3}$, the $\Upsilon$ is
positive upto $\eta=0.03$, above that it becomes negative up to $\eta=0.07$.
Above $\eta=0.07$, $\Upsilon$ is again positive for all allowed values of
$\eta$. However, at number density $n_{0}=10^{22}cm^{-3}$, the values of
$\eta$ is allowed up to $0.07$, above that it becomes negative up to
$\eta=0.18$. Above $\eta=0.18$, $\Upsilon$ is again positive for all allowed
values of $\eta$. If we further increase the density $n_{0}=10^{23}cm^{-3}$,
the range of $\eta$ is allowed up to $0.18$, above that it becomes negative.

To examine the nonlinear periodic IAWs, we get the following cnoidal wave
solution%
\begin{equation}
\phi_{1}=\left[  \phi_{2}+\left(  \phi_{0}-\phi_{2}\right)  cn^{2}\left\{
\left(  \frac{C\sqrt{\phi_{0}-\phi_{3}}}{\sqrt{12D}}\right)  \zeta,m\right\}
\right]  ,\label{18}%
\end{equation}
where $cn$ is the Jacobian elliptic function, and the parameters $m$
represents the nonlinearity strength\ is defined as%
\[
m=\sqrt{\frac{\phi_{0}-\phi_{2}}{\phi_{0}-\phi_{3}}},
\]
where $m^{2}$ is varied as $0<m\leq1$.

\section{Numerical analysis}

\subsection{Ion-acoustic Wave(IAW)}

In this section, we numerically analyze the results of IA cnoidal waves in a
magnetized quantum e-i plasma. For the numerical analysis typical parameters:
$n_{0}=10^{21}-10^{23}cm^{-3}$and $B_{0}=10^{5}-10^{7}G$ are used which are
relevant to laboratory degenerate plasma \cite{Akbari-2012}.

Fig. (3) represents the comparison of Sagdeev potentials $U(\phi_{1})$, phase
potraits and cnoidal waves under the self-consistent field approximation and
exchange-correlation potential effects. The Fig.(3a) demonstrates that the
presence of exchange correlation potential increases the amplitude of the
potential and reduce the depth.

The phase plane plot is displayed in Fig.(3b) clearly reveals that the
amplitude increases in the presence of exchange correlation potential effects.
Fig.(3c) shows that the presence of exchange-correlation effect leads to
increase the amplitude and wavelength of the cnoidal wave structure. Fig.(4)
represents the influence of spin polarization on the Sagdeev potentials
$U(\phi_{1})$, phase potraits and cnoidal waves. Fig.(4a) depicts that the
spin polarization enhances both the amplitude and width of Sagdeev potential.
In Fig.(4b) it is noted that amplitude of the cnoidal wave increases with spin
polarization. The plots of cnoidal wave structures are plotted in Fig.(4c),
which shows the enhancement of amplitude but the wavelength decreases in spin effect.

\subsection{Spin Electron- acoustic Wave(SEAW)}

As we have discussed earlier the IAW, now following the same procedure for
SEAW, we derive the KdV equation%
\begin{equation}
D\partial_{T}\phi_{1}-Z\partial_{\xi}\phi_{1}^{2}+\partial_{\xi}^{3}\phi
_{1}=0, \label{19}%
\end{equation}
and the coefficients of KdV's equation $D$ and $Z$ are given as:%
\begin{equation}
D=\frac{\sqrt{3}}{\upsilon_{Fe}r_{De}^{2}}D_{0}, \label{19a}%
\end{equation}
where%
\begin{equation}
D_{0}=\frac{w(1-\eta)}{\left(  w^{2}-(1-\eta)^{2/3}\right)  ^{2}}%
+\frac{w(1+\eta)}{\left(  w^{2}-(1+\eta)^{2/3}-\frac{\chi e^{2}n_{0e}^{1/3}%
}{mv_{Fe}^{2}}\right)  ^{2}}, \label{19b}%
\end{equation}
and%
\begin{equation}
Z=\frac{e}{m_{e}}\frac{3}{\upsilon_{Fe}^{2}r_{De}^{2}}Z_{0}, \label{20}%
\end{equation}
where%
\begin{align}
&  \left.  Z_{0}=\frac{(1-\eta)}{\left(  w^{2}-(1-\eta)^{2/3}\right)  ^{2}%
}\left(  \frac{1}{2}+\frac{w^{2}+\frac{1}{3}(1-\eta)^{2/3}}{w^{2}%
-(1-\eta)^{2/3}}\right)  \right. \nonumber\\
&  \text{ \ \ \ \ \ \ }\left.  +\frac{(1+\eta)}{\left(  w^{2}-(1+\eta
)^{2/3}-\frac{\chi e^{2}n_{0e}^{1/3}}{mv_{Fe}^{2}}\right)  ^{2}}\left(
\frac{1}{2}+\frac{w^{2}+\frac{1}{3}(1+\eta)^{2/3}-\frac{\chi e^{2}n_{0e}%
^{1/3}}{mv_{Fe}^{2}}}{w^{2}-(1+\eta)^{2/3}-\frac{\chi e^{2}n_{0e}^{1/3}%
}{mv_{Fe}^{2}}}\right)  \right. \nonumber\\
&  \text{ \ \ \ \ \ \ }\left.  +\left(  \frac{(1+\eta)}{\left(  w^{2}%
-(1+\eta)^{2/3}-\frac{\chi e^{2}n_{0e}^{1/3}}{mv_{Fe}^{2}}\right)  ^{3}%
}\right)  \left(  \frac{1}{6}\right)  \left(  \frac{\chi e^{2}n_{0e}^{1/3}%
}{mv_{Fe}^{2}}\right)  ,\right.  \label{21}%
\end{align}
The expression for the phase velocity $(w)$ of SEA wave in the normalized form
is written as:%
\begin{equation}
w=\sqrt{\frac{1}{2}\left(  1-\eta\right)  \left(  \left(  1+\eta\right)
^{2/3}-\frac{\chi e^{2}n_{0e}^{1/3}}{mv_{Fe}^{2}}\right)  +\frac{1}{2}\left(
1+\eta\right)  \left(  1-\eta\right)  ^{2/3}}. \label{22}%
\end{equation}

The Fig.(5) shows the phase velocity of SEA wave plotted against the spin
polarization $\eta$. It can be seen from the figure that under the
self-consistent field approximation the phase velocity of SEA wave is
decreases against $\eta$ and in the presence of exchange effects it further
decreases. It means the exchange effects decreases the phase velocity of SEAW.
Consider the propagation of SEA waves along the magnetic field with the
Exchange Correlation Potential and spin polarization. We derive the periodic
wave solution of SEAW and its nonlinear equations are given as:%
\begin{equation}
U\left(  \phi_{1}\right)  =-\frac{C}{6A}\phi_{1}^{3}-\frac{U_{0}}{2A}\phi
_{1}^{2}+\rho_{0}\phi_{1}, \label{23}%
\end{equation}%
\begin{equation}
\frac{d\phi_{1}}{d\zeta}=\pm\sqrt{\frac{C}{3A}\left(  \phi_{0}-\phi
_{1}\right)  \left(  \phi_{1}-\phi_{2}\right)  \left(  \phi_{1}-\phi
_{3}\right)  }, \label{24}%
\end{equation}%
\begin{equation}
\Upsilon=\frac{U_{0}^{2}}{C^{2}}+\frac{2A\rho_{0}}{C}, \label{25}%
\end{equation}%
\begin{equation}
\phi_{1}=\phi_{2}+\left(  \phi_{0}-\phi_{2}\right)  cn^{2}\left[  \frac
{C\sqrt{\phi_{0}-\phi_{3}}}{\sqrt{12A}}\zeta,m\right]  , \label{26}%
\end{equation}
where $C=2Z/D$ and $A=1/D$. Eq.\ref{25} gives the condition for existence of
cniodal wave which is plotted in Fig. (6) against $\eta$ for different values
of number density. It can be seen from Fig.(6) that at number density,
$n_{0}=10^{22}cm^{-3}$, the $\Upsilon$ is positive for all allowed range of
$\eta$. However, at number density, $n_{0}=10^{23}cm^{-3}$, the value of
$\eta$ is allowed up to $0.5$. Above that, it becomes negative. If we further
increase the density, $n_{0}=10^{24}cm^{-3}$, the allowed range of $\eta$
further decreases and now $\eta$ is allowed up to $0.025$, above that it
becomes negative again.

Fig.7(a)-(c) shows the comparison of cnoidal structures with and without
Exchange Correlation Potential. Fig.7(a,b,c) shows that Exchange Correlation
Potential effect changes the polarity of the Sagdeev potential, phase potrait
and pulse structure. It means that exchange effect are responsible to change
the structures from compressive to rarefactive.

Fig.8(a)-(c) represents the effect of spin polarization on the non-linear
structures. Fig. 8(a) shows that amplitude of the Sagdeev potential well
increases but its depth decreases with increasing value of $\eta$. Fig. 8(b),
shows that as the spin polarization increases, the size of the cnoidal wave
structures increases as well as increase in the amplitude as shown in Fig. 8(c).

\section{Conclusion}

In this study, we examined the periodic structures associated with
longitudinal parallel propagating waves, namely ion-acoustic (IA) and spin
electron-acoustic (SEA) waves, in spin-polarized plasma. We considered the
self-consistent field approximation along with the contribution of exchange
effects. By applying the reductive perturbation method, we derived the
Korteweg-de Vries (KdV) equation and its cnoidal solutions. In the
self-consistent field approximation, the phase velocity of the IA wave remains
nearly constant for all allowed values of $\eta$ (ranging from $0$ to $1$).
However, when exchange effects are included, the phase velocity of the IA wave
shows a significant change and becomes positive for $\eta>0.2$. Additionally,
it is observed that the range of $\eta$ decreases with increasing number
density ($n_{0}$), indicating that nonlinear structures do not exist for all
values of $\eta$. The cnoidal structures of the IA wave remain compressive
under both the self-consistent field approximation and in the presence of
exchange effects. For the SEA wave, the phase velocity decreases in the
presence of exchange effects compared to the self-consistent field
approximation. We also noted that cnoidal structures do not exist for all
allowed values of $\eta$; instead, they are confined to specific values of
$\eta$.

For the cnoidal structures of SEAW, the presence of exchange-correlation
potential is responsible to change the nonlinear structures from rarefactive
to compressive. Regarding the applicability of results: our results are valid
for the density $n_{0}\ll10^{25}cm^{-3}$ because exchange effect become
important and can not be ignored as compared to Fermi pressure. However, for
$n_{0}\gg10^{25}cm^{-3}$ Fermi pressure overcomes over the exchange
interaction for all polarizations. Further this model works in the
nonrelativistic limit. Additionally, this study plays a significant role in
enhancing our understanding formation of high frequency new spin dependent
waves and low frequency waves in the laboratory plasma environments.

\begin{figure}[h] 
    \centering
    \includegraphics[width=0.7\textwidth]{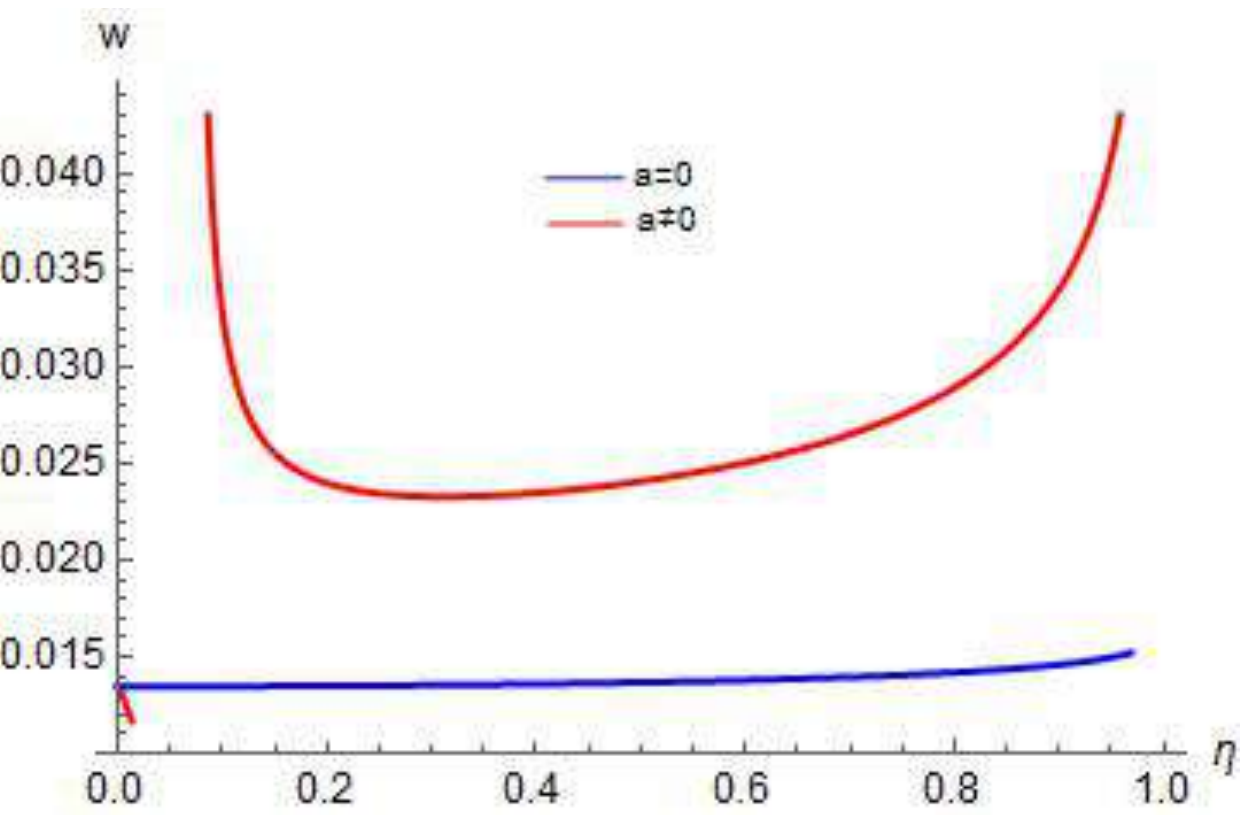} 
    \caption{Graphs of phase velocity against spin polarization $\eta$ in the
presence and absence of exchange correlation. Here $a=\frac{\chi e^{2}%
n_{0e}^{1/3}}{mv_{Fe}^{2}}$}
    \label{fig:1}
\end{figure}

\begin{figure}[h] 
    \centering
    \includegraphics[width=0.7\textwidth]{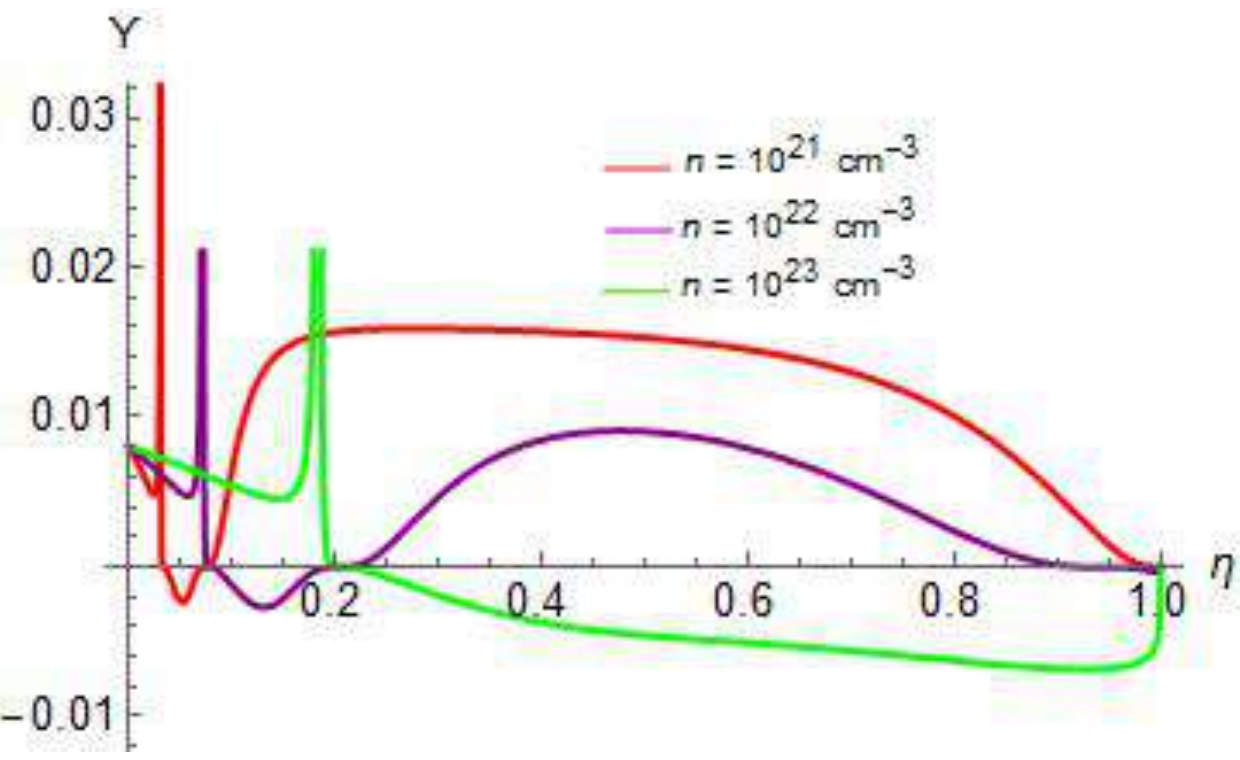} 
    \caption{Graphs of $\Upsilon$ eq.(17) against spin polarization $\eta$ in the presence of exchange correlation with the variation of density.}
    \label{fig:2}
\end{figure}

\begin{figure}[h] 
    \centering
    \includegraphics[width=0.7\textwidth]{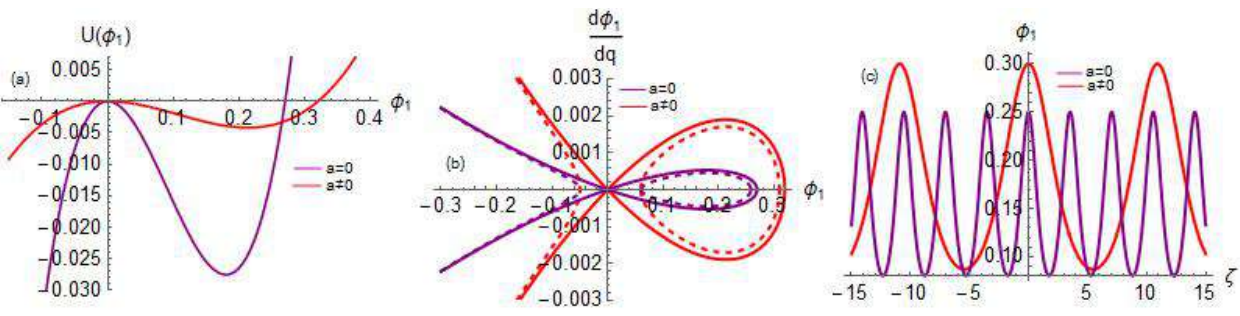} 
    \caption{Shows the comparison of Sagdeev potential $U(\phi_{1})$, phase plane
plots and cnoidal wave structures with and without exchange. Here $\phi_{1}$
is normalized with $e/\varepsilon_{Fe}$ and space variable $\zeta$ is
normalized with $\lambda_{De}$.}
    \label{fig:3}
\end{figure}

\begin{figure}[h] 
    \centering
    \includegraphics[width=0.7\textwidth]{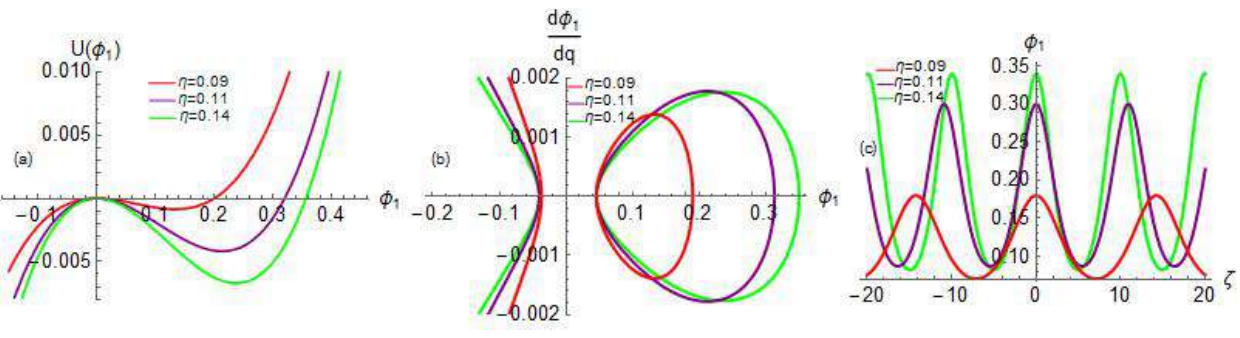} 
    \caption{shows the effect of spin polarization $\eta$ on Sagdeev potential, phase
plane plots and cnoidal wave structures.}
    \label{fig:4}
\end{figure}

\begin{figure}[h] 
    \centering
    \includegraphics[width=0.7\textwidth]{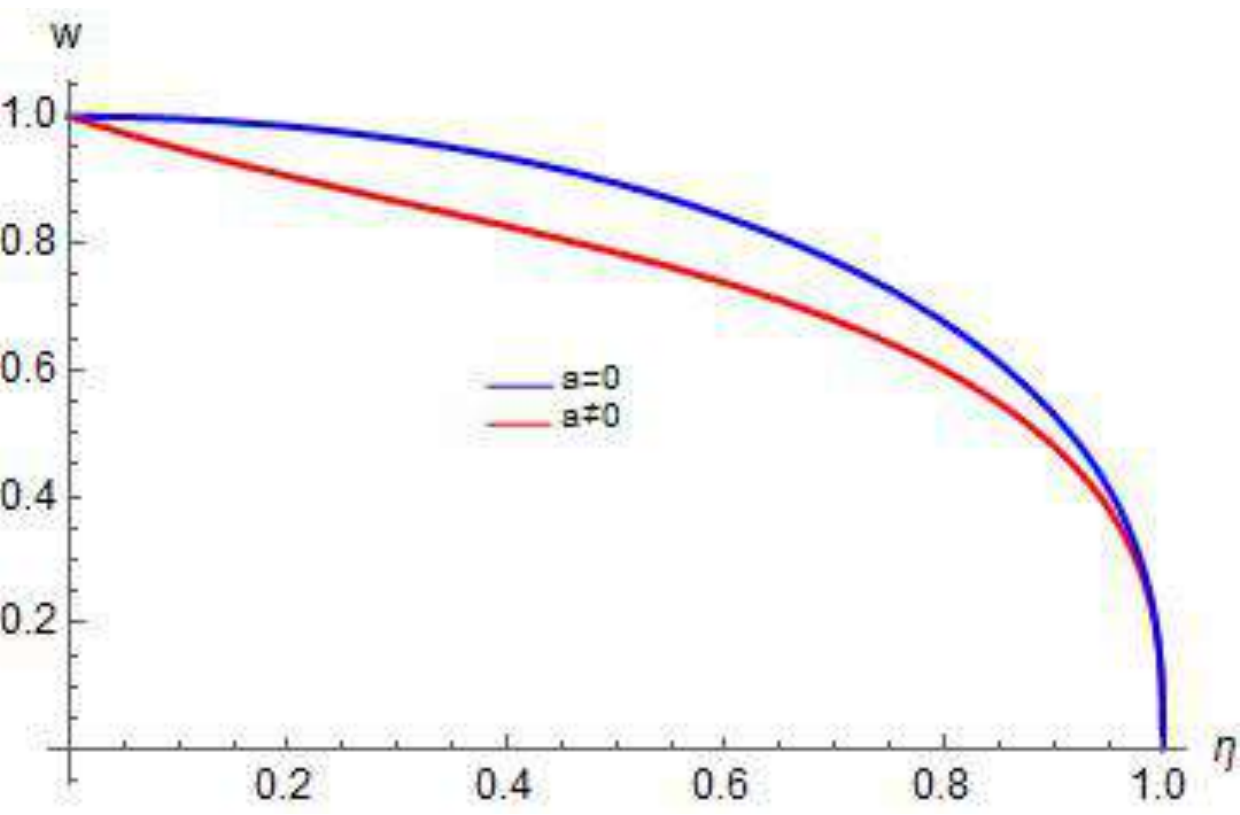} 
    \caption{shows the graphs of phase velocity against spin polarization $\eta$ in
the presence and absence of exchange correlation. Here the density is in
$cm^{-3}$.}
    \label{fig:5}
\end{figure}

\begin{figure}[h] 
    \centering
    \includegraphics[width=0.7\textwidth]{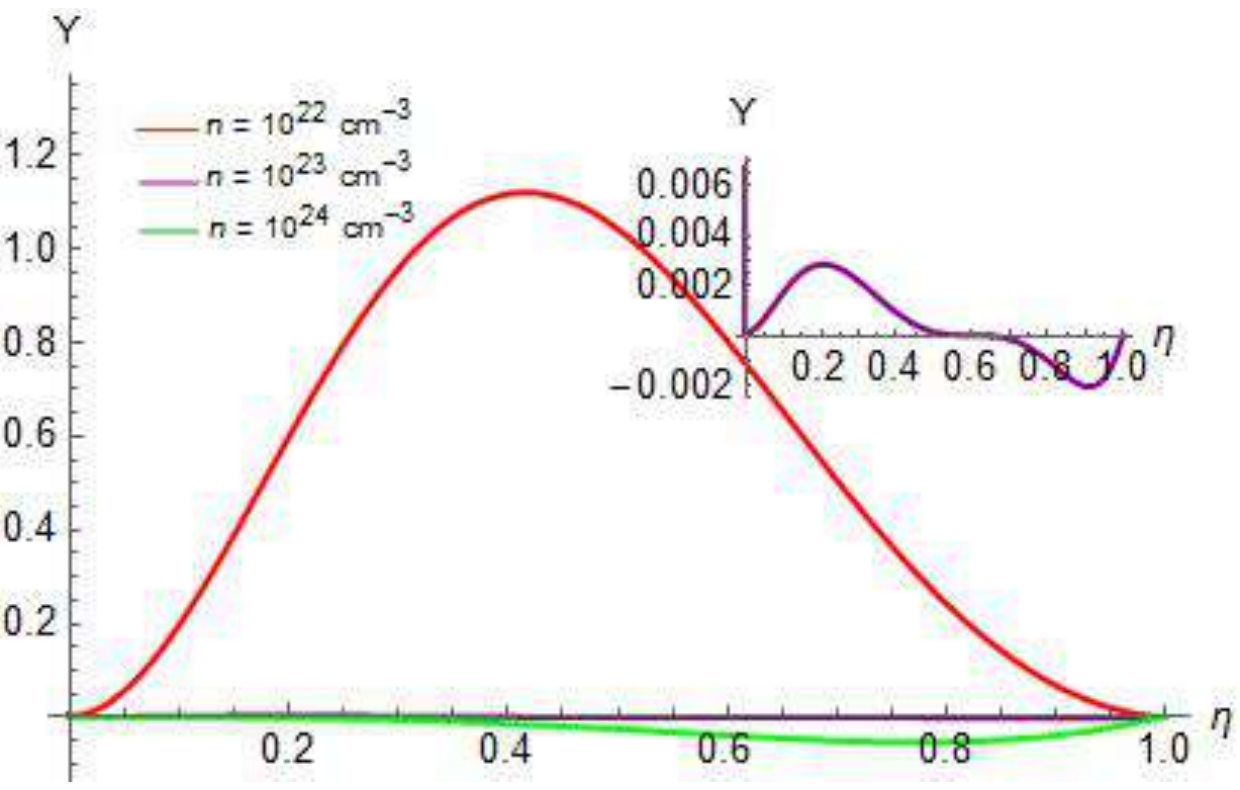} 
    \caption{shows the variation of $\Upsilon$ against spin polarization $\eta$.}
    \label{fig:6}
\end{figure}

\begin{figure}[h] 
    \centering
    \includegraphics[width=0.7\textwidth]{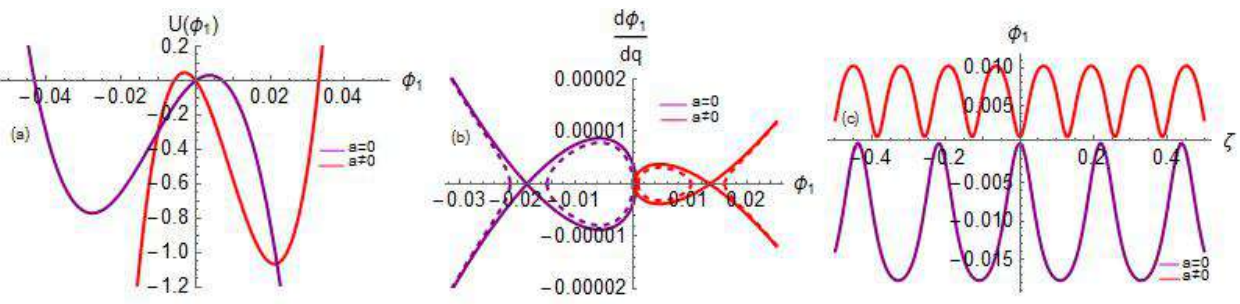} 
    \caption{Graphs of nonlinear spin electron-acoustic wave in the absence and
presence of exchange correlation are plotted at fixed values of equilibrium
density ($n_{0}=10^{23}cm^{-3}$) and magnetic field ($B_{0}=10^{7}G$). Here
$\phi_{1}$ is normalized with $e/\varepsilon_{Fe}$ and space variable $\zeta$
is normalized with $\lambda_{De}$.}
    \label{fig:7}
\end{figure}

\begin{figure}[h] 
    \centering
    \includegraphics[width=0.7\textwidth]{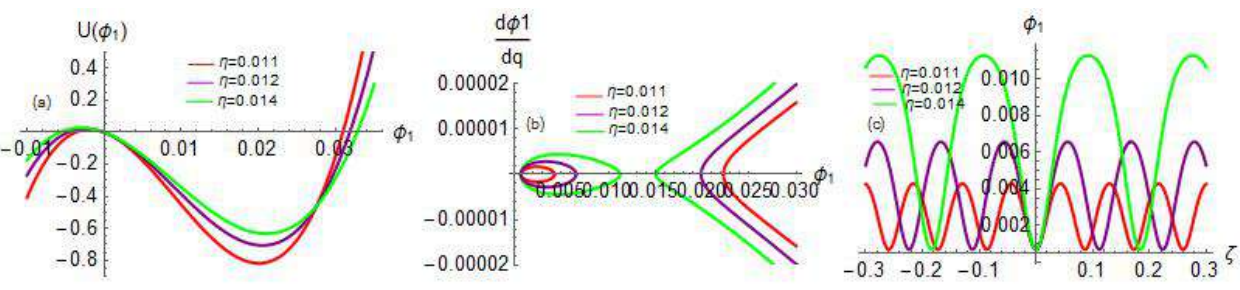} 
    \caption{Graphs of nonlinear SEA wave in are plotted at fixed value of electron
density, i.e., $n_{0}=10^{23}cm^{-3}$ and different values of magnetic field
to see the effect of spin polarization $\eta$.}
    \label{fig:8}
\end{figure} 

\end{document}